\documentclass[12pt,a4paper]{article} 

\usepackage[utf8]{inputenc}
\usepackage[T1]{fontenc}
\usepackage{lmodern} 
\usepackage{geometry} 
\usepackage{graphicx}
\usepackage{multirow}
\usepackage{amsmath,amssymb,amsfonts}
\usepackage{amsthm}
\usepackage{mathrsfs}
\usepackage[title]{appendix}
\usepackage{xcolor}
\usepackage{textcomp}
\usepackage{manyfoot}
\usepackage{booktabs}
\usepackage{algorithm}
\usepackage{algorithmicx}
\usepackage{algpseudocode}
\usepackage{listings}
\usepackage{subcaption,caption}
\usepackage{epstopdf}
\usepackage{url}
\usepackage{rotating}
\usepackage{placeins}
\usepackage{float}
\usepackage{enumerate}
\usepackage{setspace}
\usepackage{tikz}
\usepackage{cite}
\usepackage[percent]{overpic}
\usepackage[colorlinks=true, linkcolor=blue, citecolor=red, urlcolor=magenta]{hyperref}

\newtheoremstyle{thmstyleone}%
  {3pt}{3pt}
  {\itshape}
  {}
  {\bfseries}
  {.}
  { }
  {}

\newtheoremstyle{thmstyletwo}%
  {3pt}{3pt}%
  {\normalfont}%
  {}%
  {\bfseries}%
  {.}%
  { }%
  {}%

\newtheoremstyle{thmstylethree}%
  {3pt}{3pt}%
  {\normalfont}%
  {}%
  {\bfseries}%
  {.}%
  { }%
  {\itshape}%

\theoremstyle{thmstyleone}

\theoremstyle{thmstyletwo}

\theoremstyle{thmstylethree}
\newtheorem{definition}{Definition}

\begin{document}

\title{Emergence of fractal structures from breather interactions in the $(2+1)$-dimensional Konopelchenko--Dubrovsky equation}

\author{
Snehalata Nasipuri\textsuperscript{1}, Prasanta Chatterjee\textsuperscript{2}, Saugata Dutta\textsuperscript{2}\thanks{Email: \href{mailto:saugatadutta.apd@gmail.com}{saugatadutta.apd@gmail.com}}\\[4pt]
\textsuperscript{1}\small Department of Mathematics, Sikkim Manipal Institute of Technology, Majitar-737136, India\\
\textsuperscript{2}\small Department of Mathematics, Siksha Bhavana, Visva-Bharati, Santiniketan-731235, India}
\date{}
\maketitle
\begin{abstract}
 Fractal structures generated through nonlinear breather interactions are investigated for the $(2+1)$-dimensional Konopelchenko--Dubrovsky (KD) equation by means of the Hirota bilinear method. The bilinear form of the system is first derived, after which breather interaction solutions are constructed analytically through suitable auxiliary functions. It is shown that the interaction of breather waves in the coupled nonlinear environment gives rise to highly intricate multiscale patterns exhibiting self-similar behaviour under successive magnification. To characterize the geometric complexity of the obtained structures, a three-dimensional voxel-based box-counting method is employed. The computed dimensions are found to be non-integer, confirming the fractal nature of the generated patterns. In addition, relative error analysis, standard error estimation, bootstrap standard deviation and convergence analysis are performed to examine the robustness and reproducibility of the estimated dimensions. The present work suggests that nonlinear breather interactions in coupled dispersive systems may provide a natural mechanism for the emergence of fractal geometries and complex multiscale structures. The combined analytical and quantitative framework developed here may provide further insight into nonlinear energy localization and scale-dependent structures arising in fluid dynamics, plasma physics and nonlinear wave propagation.
\end{abstract}




\maketitle

\section{Introduction}

Nonlinear partial differential equations (NLPDEs) have long played an important role in the mathematical modelling of complex physical phenomena arising in fluid dynamics, plasma physics, nonlinear optics, condensed matter systems and wave propagation theory. In recent years, coupled nonlinear systems have received increasing attention because they describe the interaction between multiple physical components or wave fields. Such coupled interactions often generate dynamical behaviours that are significantly richer and more complicated than those observed in single nonlinear equations. In fluid mechanics, for example, coupled systems are frequently associated with multilayer fluid flows, interacting shallow-water waves and stratified media, where continuous energy exchange between different layers may produce highly intricate wave structures. Similar coupled interactions also arise in plasmas and optical systems, where nonlinear energy transfer between different modes can lead to the formation of localized patterns, wave instabilities and multiscale structures.

Alongside the study of nonlinear waves, specially the investigation of fractal structures has emerged as an active topic in modern nonlinear science. Fractal patterns are generally characterized by self-similar geometries that persist across different spatial scales and are commonly associated with non-integer dimensions. Such structures appear naturally in many physical systems, including turbulent flows, plasma transport, optical filamentation and complex surface growth phenomena. The presence of fractal geometry is often related to nonlinear energy cascades, irregular wave interactions and scale-dependent pattern formation. Consequently, the study of fractal behaviour in nonlinear evolution equations has become increasingly important for understanding multiscale dynamics in complex physical environments.

Compared with single nonlinear equations, coupled NLPDEs provide a more suitable framework for the generation of complicated geometrical structures because the interaction between different wave components allows additional mechanisms of energy redistribution and nonlinear modulation. For instance, in multilayer fluid systems, structures generated in one layer may continuously exchange energy with structures in another layer, resulting in more complicated collective dynamics. Such interaction processes may naturally produce recursive and scale-dependent patterns that resemble fractal geometries. Therefore, the investigation of fractal structures in coupled nonlinear systems is not only mathematically interesting but also physically relevant for understanding complex nonlinear processes in interacting media.

Several recent studies have demonstrated that nonlinear dispersive systems can exhibit self-similar waveforms and fractal geometries through suitable analytical constructions. In many of these works, fractal structures were generated by imposing specially selected auxiliary functions, such as trigonometric, logarithmic or Jacobi elliptic functions, into previously derived analytical solutions obtained through Riccati-type transformations \cite{Chatterjee2025,Dutta2025,Dutta2026}. Although these approaches successfully produced recursive multiscale patterns, the emergence of fractal structures was mainly dependent on the choice of imposed functional forms. As a result, a more systematic mechanism responsible for the generation of fractal structures in coupled nonlinear systems has remained unclear.

Motivated by this observation, the present work investigates whether fractal structures can emerge directly from nonlinear breather interactions constructed through the Hirota bilinear method (HBM). The HBM is one of the most powerful and systematic analytical techniques for nonlinear evolution equations and has been widely used to derive multisoliton, lump, rogue-wave and breather solutions. Since breather waves represent localized oscillatory structures with strong nonlinear interaction properties, it is expected that their interaction in coupled environments may naturally generate highly intricate multiscale geometries. From this viewpoint, the present study suggests that fractal structures may arise as a consequence of nonlinear breather interactions in coupled systems rather than only through externally imposed auxiliary functional forms. In a recent related study, we observed that breather interactions in coupled Boussinesq-type systems can also generate fractal-like structures \cite{epl}, further supporting the idea that nonlinear breather dynamics may provide a natural mechanism for the emergence of multiscale geometries in coupled nonlinear media.

In this work, fractal structures are investigated for the $(2+1)$-dimensional Konopelchenko--Dubrovsky (KD) equation
\begin{equation}\label{eq1}
\begin{split}
& u_t-u_{xxx}-6 b u u_x + \frac{3}{2} a^2 u^2 u_x -3 v_y +3 a u_x v=0, \\
& u_y=v_x,
\end{split}
\end{equation}
where $a$ and $b$ are arbitrary constants, while $u(x,y,t)$ and $v(x,y,t)$ denote the amplitudes of the interacting waves. The KD equation, introduced by Konopelchenko and Dubrovsky \cite{KONOPELCHENKO198415}, is an important nonlinear model for studying wave propagation, interaction and modulation in several physical contexts. In fluid dynamics, the equation describes nonlinear dispersive wave interactions, while in solid mechanics it models certain stress--strain dynamics in elastic media. The equation also appears in nonlinear optics, where it is related to nonlinear light propagation, self-focusing and optical soliton dynamics.

Because of its rich mathematical structure and physical importance, the KD equation has been studied through several analytical approaches. Non-traveling wave solutions were obtained using the improved tanh-function method \cite{SHENG20061213}, while Jacobi elliptic function solutions were derived through the modified $F$-expansion method \cite{WANG2010216}. Multiple lump solutions were investigated by means of the Hirota bilinear method \cite{Xu_2011} and  soliton solutions expressed in Gram determinant form were constructed through Sato theory and Hirota techniques \cite{yuan_18}. Breather waves, Wronskian solutions and Painlev\'e properties of the system have also been examined in several works \cite{gu_23,xu_11,Xu_2011}. More recently, rogue waves and Akhmediev breathers on elliptic backgrounds have been investigated in related nonlinear wave equations \cite{grin_18,feng_20}.

To the best of our knowledge, fractal structures generated through breather interactions in the KD equation have not yet been reported. The primary objective of the present work is therefore to construct analytical breather interaction solutions through the Hirota bilinear framework and to investigate the emergence of self-similar multiscale patterns arising from these interactions. Particular attention is given to the possibility that coupled nonlinear environments may naturally support fractal-type geometries through nonlinear energy exchange mechanisms between interacting breather modes.

To quantify the geometric complexity of the obtained structures, a three-dimensional voxel-based box-counting method is employed. The resulting dimensions are found to be non-integer, supporting the fractal nature of the observed patterns. Furthermore, statistical diagnostics including relative error analysis, standard error estimation, bootstrap standard deviation and convergence analysis are incorporated to examine the robustness and reproducibility of the estimated fractal dimensions. The combined analytical construction, graphical visualization and quantitative fractal analysis provide a systematic framework for characterizing multiscale geometries generated by nonlinear breather interactions in higher-dimensional coupled systems.

The remainder of the paper is organized as follows. In Sect.~2, the Hirota bilinear form of the KD equation and the corresponding breather interaction solutions are derived. Sect.~3 is devoted to the graphical investigation of the self-similar structures arising from the obtained solutions. In Sect.~4, fractal dimensions and their statistical validation are discussed through voxel-based box-counting analysis. Finally, the main conclusions of the present work are summarized in Sect.~5.

\section{Bilinear formulation and breather interaction solutions}\label{frac}

The $(2+1)$-dimensional KD equation \eqref{eq1} is first converted into its bilinear form using the Hirota bilinear method. This formulation provides a convenient framework for constructing analytical solutions that describe nonlinear wave interactions.
To this end, the dependent variable transformations
\begin{equation}\label{tran1}
\begin{split}
u(x,y,t) &= \frac{2}{a} \left[\ln \frac{g(x,y,t)}{f(x,y,t)} \right]_x, \\
v(x,y,t) &= \frac{2}{a} \left[\ln \frac{g(x,y,t)}{f(x,y,t)} \right]_y
\end{split}
\end{equation}
are introduced, where $g(x,y,t)$ and $f(x,y,t)$ are auxiliary functions.

The Hirota bilinear operator is defined as
\begin{equation}\label{def}
\begin{aligned}
\mathcal{D}_x^n \mathcal{D}_y^p \mathcal{D}_t^q (g \cdot f)
= \left(\partial_x - \partial_{x'}\right)^n
\left(\partial_y - \partial_{y'}\right)^p
\left(\partial_t - \partial_{t'}\right)^q
\, g(x,y,t) f(x',y',t') \big|_{x=x',y=y',t=t'}.
\end{aligned}
\end{equation}

Under the transformation \eqref{tran1}, Eq. \eqref{eq1} reduces to the bilinear system
\begin{equation}\label{biform}
\begin{split}
\left(\mathcal{D}_t - \mathcal{D}_x^3 + 3 \mathcal{D}_x \mathcal{D}_y + 6A \mathcal{D}_y \right)(g \cdot f) &= 0, \\
\left(\mathcal{D}_y + \mathcal{D}_x^2 + 2A \mathcal{D}_x \right)(g \cdot f) &= 0,
\end{split}
\end{equation}
where $A=-b/a$.
To capture interaction dynamics, the auxiliary functions are chosen as a combination of oscillatory and localized components,
\begin{equation}\label{exp}
\begin{split}
g(x,y,t) &= a_1 \cos(\theta_1) + a_2 \cosh(\theta_2), \\
f(x,y,t) &= b_1 \cos(\theta_1) + b_2 \cosh(\theta_2), \quad \textnormal{where} \quad
\theta_1 = p_1 x + q_1 y + l_1 t, \qquad
\theta_2 = p_2 x + q_2 y + l_2 t.
\end{split}
\end{equation}
Substitution of \eqref{exp} into \eqref{biform} and separation of the independent functional terms lead to a set of algebraic constraints. These are satisfied provided that
\begin{equation}\label{sol1}
\begin{split}
a_1 &= -\frac{i a_2 p_2}{p_1}, \quad
b_1 = \frac{i b_2 p_2}{p_1}, \quad q_1 = -2A p_1, \quad
q_2 = -2A p_2, \\
l_1 &= -p_1(-12A^2 + p_1^2 - 3p_2^2), \quad l_2 = p_2(12A^2 - 3p_1^2 + p_2^2).
\end{split}
\end{equation}

With these relations, explicit analytical expressions for $u(x,y,t)$ and $v(x,y,t)$ are obtained from \eqref{tran1}. The resulting solutions describe the nonlinear interaction between two breather-type modes, arising from the coupling of trigonometric and hyperbolic structures.

The expressions for $u(x,y,t)$ and $v(x,y,t)$ are given by
\begin{equation}\label{solu}
\begin{split}	
&u(x,y,t)=\Bigg[4 i p_1 p_2 \Bigg(p_1 \cosh\Big[
	p_2 (x + (2 y b )/ a + 
	t (-3 p_1^2 + p_2^2 + (12 b^2)/ a^2))\Big] \\
	& \sin\Big[
	p_1 (-p_1^2 t + 3 p_2^2 t + x + (2 y b)/ a + (
	12 t b^2)/ a^2)\Big] + 
	p_2 \cos\Big[p_1 (-p_1^2 t + 3 p_2^2 t + x + (2 y b)/ a + (
	12 t b^2)/ a^2)\Big] \\
	& \sinh\Big[
	p_2 (x + (2 y b)/ a + 
	t (-3 p_1^2 + p_2^2 + (
	12 b^2)/ a^2))\Big] \Bigg) \Bigg] 	 /
	\Bigg[a \Bigg(p_2^2 \Big(\cos\Big[
	p_1 (-p_1^2 t + 3 p_2^2 t + x  \\ & + (2 y b)/ a + (
	12 t b^2)/ a^2)\Big] \Big)^2  + 
	p_1^2 \Big(\cosh\Big[
	p_2 (x + (2 y b)/ a + 
	t (-3 p_1^2 + p_2^2 + (12 b^2)/ a^2))\Big]\Big)^2 \Bigg) \Bigg]
\end{split} 
\end{equation}
\begin{equation}\label{solv}
\begin{split}
& v(x,y,t)= \Bigg[8 i p_1 p_2 b \Bigg(p_1 cosh\Big[
p_2 (x + (2 y b)/ a + 
t (-3 p_1^2 + p_2^2 + (12 b^2)/ a^2)) \Big] \\
& \sin\Big[
p_1 (-p_1^2 t + 3 p_2^2 t + x + (2 y b)/ a + (
12 t b^2)/ a^2)\Big] + 
p_2 \cos\Big[p_1 (-p_1^2 t + 3 p_2^2 t + x + (2 y b)/ a + (
12 t b^2)/ a^2)\Big] \\
& \sinh\Big[
p_2 (x + (2 y b)/ a + 
t (-3 p_1^2 + p_2^2 + (
12 b^2)/ a^2)) \Big] \Bigg) \Bigg] / \Bigg[a^2 \Bigg(p_2^2 \Big(\cos\Big[
p_1 (-p_1^2 t + 3 p_2^2 t  + x \\ &  + (2 y b)/ a + (
12 t b^2)/ a^2)\Big] \Big)^2 + 
p_1^2 \Big(\cosh\Big[
p_2 (x + (2 y b)/ a + 
t (-3 p_1^2 + p_2^2 + (12 b^2)/ a^2))\Big]\Big)^2 \Bigg) \Bigg]
\end{split} 
\end{equation}


\section{Emergence of self-similar structures in breather interactions}
In this section, the analytical breather interaction solutions, obtained using the HBM, are examined through graphical analysis. The solutions are explored under different parameter settings and observation scales in order to understand the spatial structures generated by the interaction between the underlying wave components.
The patterns are allowed to emerge directly from the interaction between the trigonometric and hyperbolic modes present in the analytical expressions. It is found that, for certain parameter regimes, the resulting profiles develop highly intricate features, where similar patterns reappear under successive magnifications.\par
In particular, Figs. 1 and 2 illustrate the evolution of the real part of $u(x,y,t)$ under successive magnifications for $a=1$, $b=0.2$, $a_2=0.1$, $b_2=-0.02$, $p_1=0.4$ and  $p_2=0.9$ fixing $y=20$ and $t=20$ respectively. It is seen that as the observation window is reduced, similar features continue to appear, while finer details become visible. This behavior suggests that the interaction between the two modes leads to patterns that persist across different scales.  A similar behavior is observed for real part of $v(x,y,t)$ in Fig. 3, showing repeated patterns under successive zooming for the same parameter scale at $t=20$. These results suggest that the breather–breather interaction leads to complex spatial organizations that exhibit scale-dependent repetition. \par 
In addition to these patterns, Fig. 4 presents the absolute and imaginary parts of the solutions \eqref{solu} and \eqref{solv}. For parameters $a=1$, $b=0.2$, $a_2=0.1$ and $p_1=0.4$, $|u(x,t)|$ (\figurename~(a)) exhibits a sharp localized peak indicative of a bright soliton formed via breather interaction, while $\operatorname{Im}[u]$ (\figurename~(a))  displays oscillatory multi-peaked behavior. In the spatial domain ( (c) and (d)), with $b_2=-0.02$, $p_2=0.9$ and  $y=20$, $|v(x,y)|$ and $\operatorname{Im}[v]$ reveal asymmetric periodic hump-like profiles with sharp ridges, illustrating energy redistribution and coupling effects between the components. These structures demonstrate the generation of stable localized waves through the balance of dispersion, nonlinearity and  cross-phase modulation. \par 
Overall, the breather interaction solutions obtained in this work generate a rich variety of spatial structures, ranging from localized soliton-like formations to intricate self-similar patterns. A quantitative characterization of the self-similarity using dimension analysis is presented in the next section. Collectively, these results reveal that the nonlinear interplay between the trigonometric and hyperbolic components naturally produces multiscale structures, with similar features recurring across different length scales. This behavior underscores the emergence of robust self-similar patterns inherent to the breather interactions in the system.

\begin{figure}[h]
    \centering
   \begin{subfigure}{0.32\textwidth}
        \includegraphics[width=\linewidth]{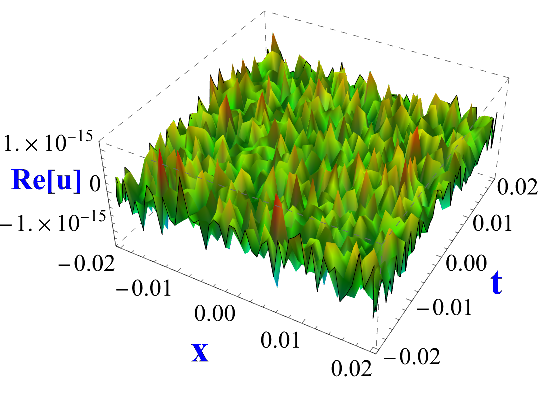}
        \caption{}
        \label{f1a}
    \end{subfigure}\hfill
    \begin{subfigure}{0.34\textwidth}
        \centering
        \includegraphics[width=\linewidth]{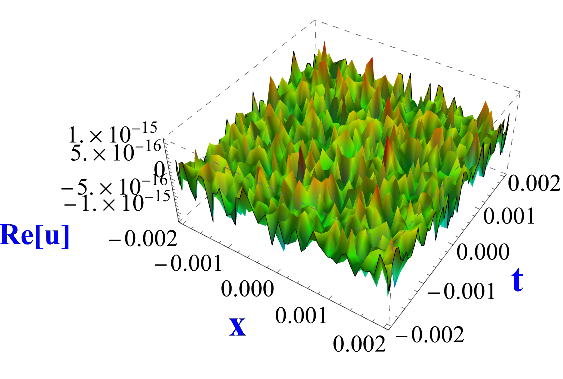}
        \caption{}
        \label{f1b}
    \end{subfigure}\hfill
    \begin{subfigure}{0.32\textwidth}
        \centering
        \includegraphics[width=\linewidth]{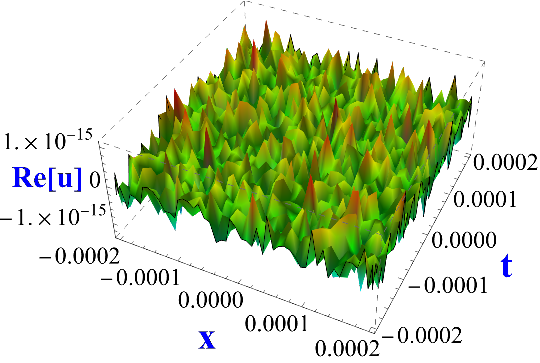}
        \caption{}
        \label{f1c}
    \end{subfigure}
    \caption{Fractal structures corresponding to the solution \eqref{solu} by fixing $y=20$ and considering the parameter values $a=1$, $b=0.2$, $a_2=0.1$, $b_2=-0.02$, $p_1=0.4$ and  $p_2=0.9$.}
    \label{f1}
\end{figure}

\begin{figure}[h]
    \begin{subfigure}[b]{0.32\textwidth}
        \includegraphics[width=\linewidth]{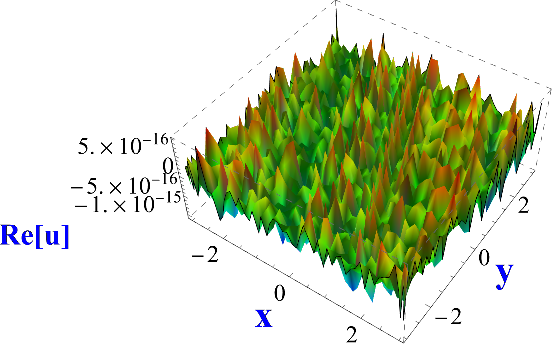}
        \caption{}
        \label{f2a}
    \end{subfigure}
    \hfill
    \begin{subfigure}[b]{0.32\textwidth}
        \includegraphics[width=\linewidth]{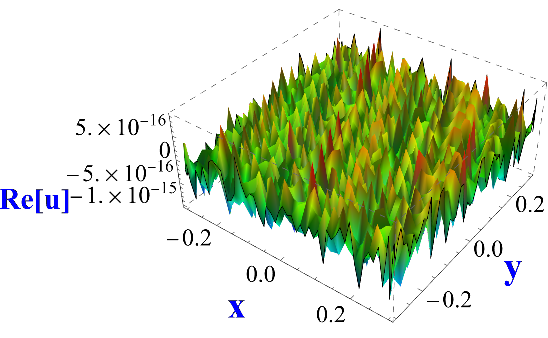}
        \caption{}
        \label{f2b}
    \end{subfigure}
    \hfill
    \begin{subfigure}[b]{0.32\textwidth}
        \includegraphics[width=\linewidth]{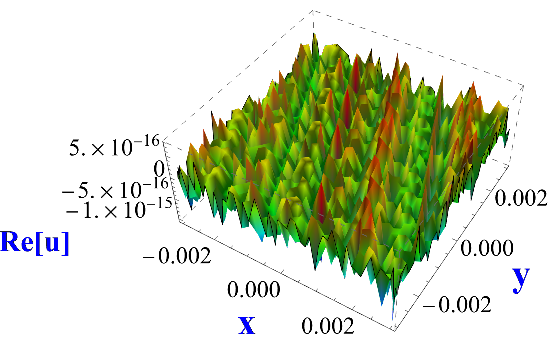}
        \caption{}
        \label{f2c}
    \end{subfigure}
    \caption{Fractal structures corresponding to the solution \eqref{solu} by fixing $t=20$ and considering the parameter values $a=1$, $b=0.2$, $a_2=0.1$, $b_2=-0.02$, $p_1=0.4$ and  $p_2=0.9$.  }
    \label{f2}
\end{figure}

\begin{figure}[h]
    \centering
    \begin{subfigure}[b]{0.32\textwidth}
        \includegraphics[width=\linewidth]{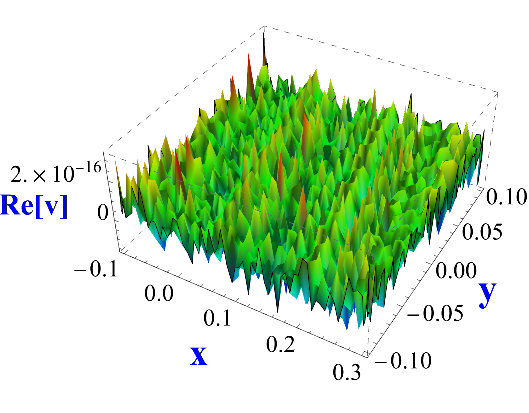}
        \caption{}
        \label{f3a}
    \end{subfigure}
    \hfill
    \begin{subfigure}[b]{0.32\textwidth}
        \includegraphics[width=\linewidth]{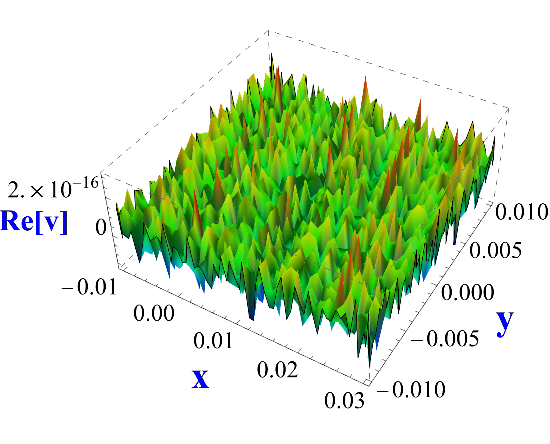}
        \caption{}
        \label{f3b}
    \end{subfigure}
    \begin{subfigure}[b]{0.32\textwidth}
        \includegraphics[width=\linewidth]{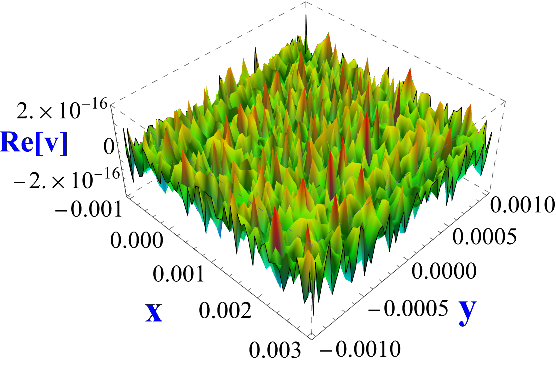}
        \caption{}
        \label{f3c}
    \end{subfigure}
    \caption{Fractal structures corresponding to the solution \eqref{solv} by considering the parameter values $a=1$, $b=0.2$, $a_2=0.1$, $b_2=-0.02$, $p_1=0.4$, $p_2=0.9$ and  $t=20$.}
    \label{f3}
\end{figure}
\FloatBarrier

\section{Fractal dimension and statistical validation}\label{S4}
The self-similar patterns observed under successive magnifications suggest that the obtained breather interaction structures possess fractal characteristics. Another important property of fractal geometry is the presence of a non-integer dimension. To quantitatively characterize the complexity of the generated structures, the box-counting dimension is employed.

\begin{definition}
Let $A$ be a bounded subset of a metric space. For a given box size $\varepsilon>0$, let $N(\varepsilon)$ denote the minimum number of boxes required to cover the set $A$. The box-counting dimension of $A$ is defined by
\[
\dim_B(A)=\lim_{\varepsilon \to 0}\frac{\log N(\varepsilon)}{\log(1/\varepsilon)},
\]
provided the limit exists.

In practical computations, the dimension is estimated by evaluating $N(\varepsilon)$ for different values of $\varepsilon$ and determining the slope of the corresponding $\log N(\varepsilon)$ versus $\log(1/\varepsilon)$ plot. The box-counting dimension measures the scaling behavior and geometrical complexity of a structure. For regular geometrical objects, the dimension is usually an integer, whereas fractal structures generally exhibit non-integer dimensions. Therefore, the appearance of a fractional dimension provides quantitative evidence of multiscale and self-similar behavior.
\end{definition}

Since the estimation of the fractal dimension involves a large number of sampled data points, it is necessary to examine the reliability and stability of the computed results. For this purpose, several statistical measures are considered. The relative error ($RE$) is used to measure the variation between successive dimension estimates and provides information about numerical consistency. The standard error ($SE$) evaluates the statistical precision of the estimated dimension with respect to the sample size. In addition, bootstrap resampling is employed to compute the bootstrap standard deviation ($\sigma_{\mathrm{boot}}$), which provides a robust estimate of uncertainty through repeated random sampling of the data.

Furthermore, the convergence measure ($|\delta_D|$) is examined by refining the box sizes and increasing the number of sampled points. If the estimated dimension remains stable under these refinements, the obtained value can be regarded as numerically reliable and independent of discretization effects.

Together, these statistical indicators provide quantitative validation for the inferred fractal dimensions and confirm the robustness of the observed multiscale structures generated by the breather interaction solutions of the $(2+1)$-dimensional KD equation.
\begin{figure}[!htbp]
    \centering
    \begin{subfigure}[b]{0.32\textwidth}
        \includegraphics[width=\linewidth]{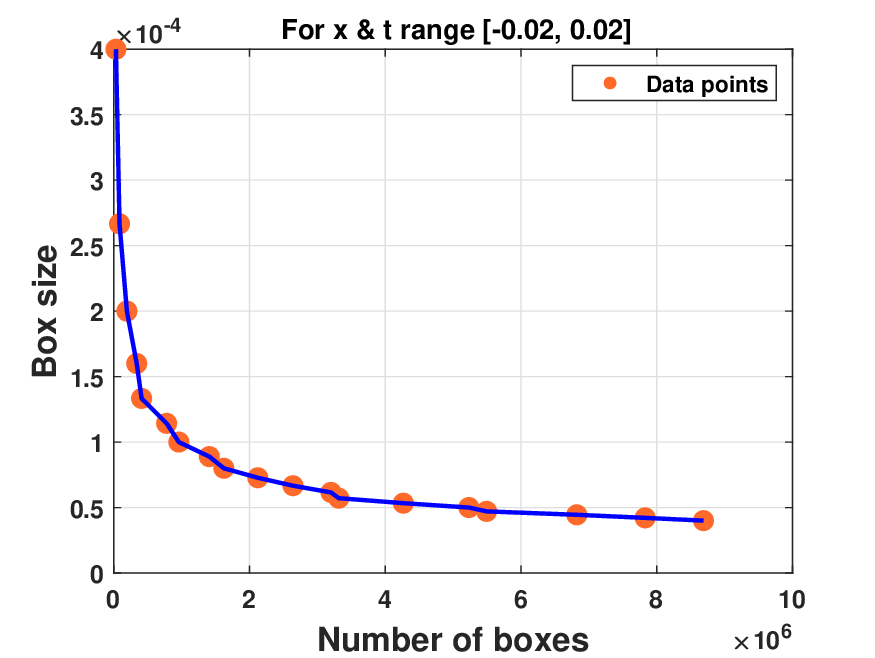}
        \caption{}
    \end{subfigure}
    \begin{subfigure}[b]{0.32\textwidth}
        \includegraphics[width=\linewidth]{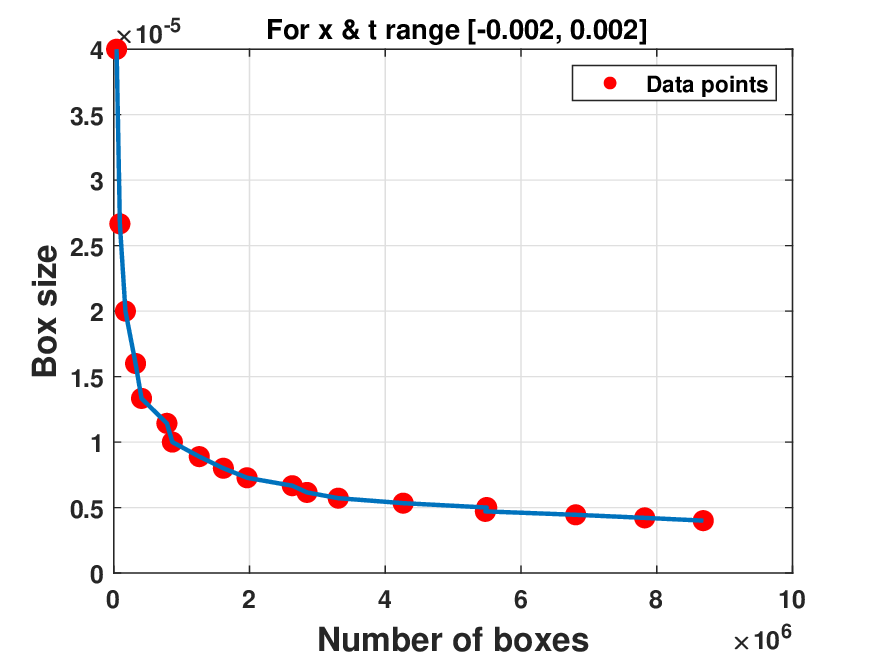}
        \caption{}
    \end{subfigure}
    \begin{subfigure}[b]{0.32\textwidth}
        \includegraphics[width=\linewidth]{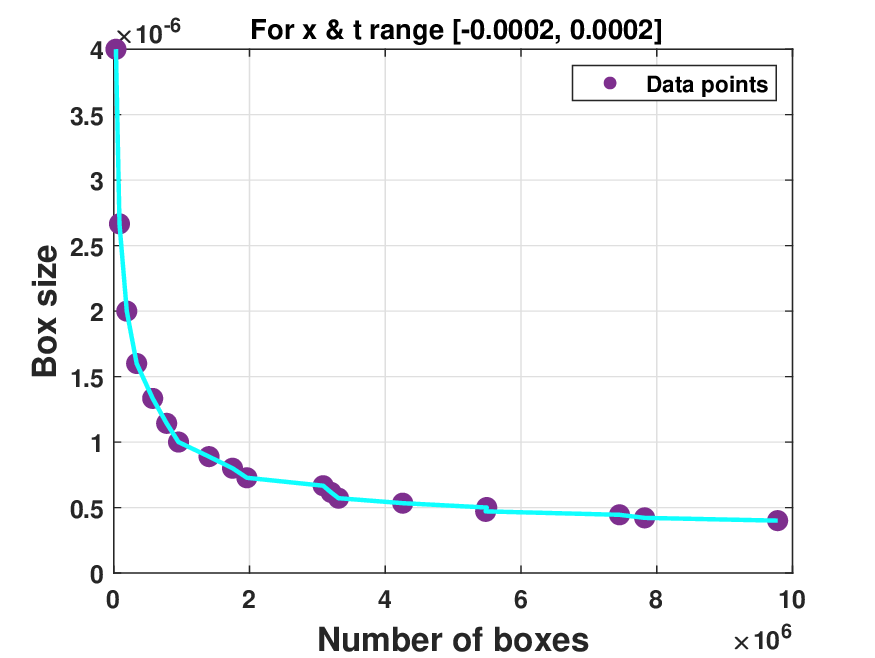}
        \caption{}
    \end{subfigure}

    \caption{Two dimensional plots of box size $(\varepsilon)$ versus number of boxes $(N(\varepsilon))$ required to cover the fractal surface corresponding to \figurename~\ref{f1} at three different zoom levels in the domain \( \xi \) and \( \eta \). Data points represent computed values from the box-counting process.}
    \label{f4}
\end{figure}
\begin{figure}[!htbp]
    \centering
    \begin{subfigure}[b]{0.32\textwidth}
        \includegraphics[width=\linewidth]{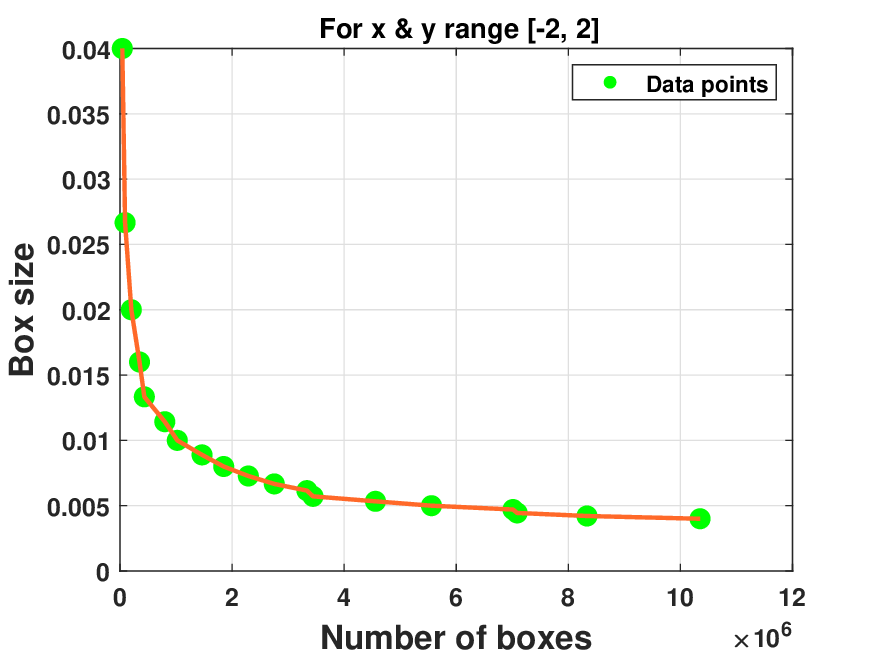}
        \caption{}
    \end{subfigure}
    \begin{subfigure}[b]{0.32\textwidth}
        \includegraphics[width=\linewidth]{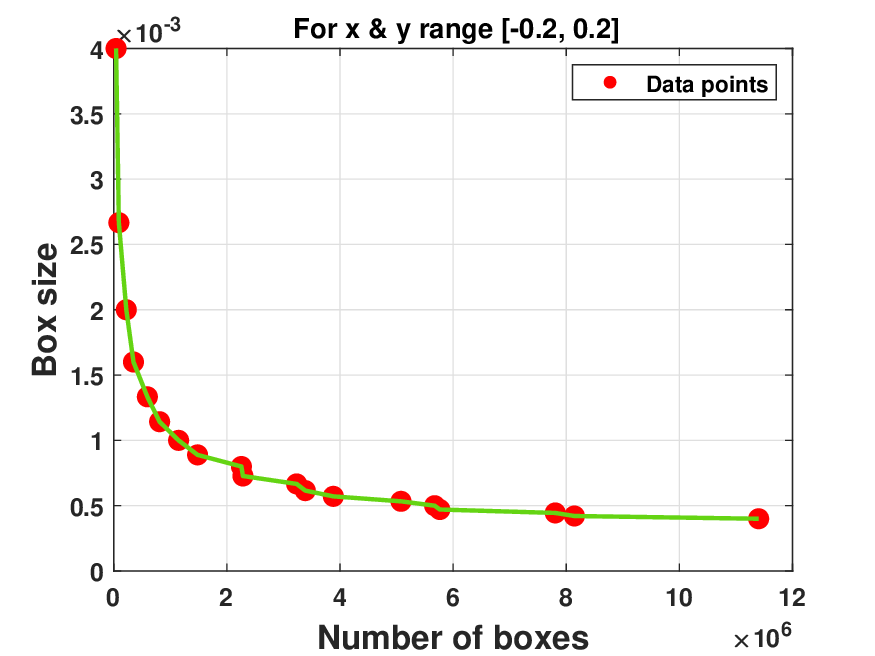}
        \caption{}
    \end{subfigure}
    \begin{subfigure}[b]{0.32\textwidth}
        \includegraphics[width=\linewidth]{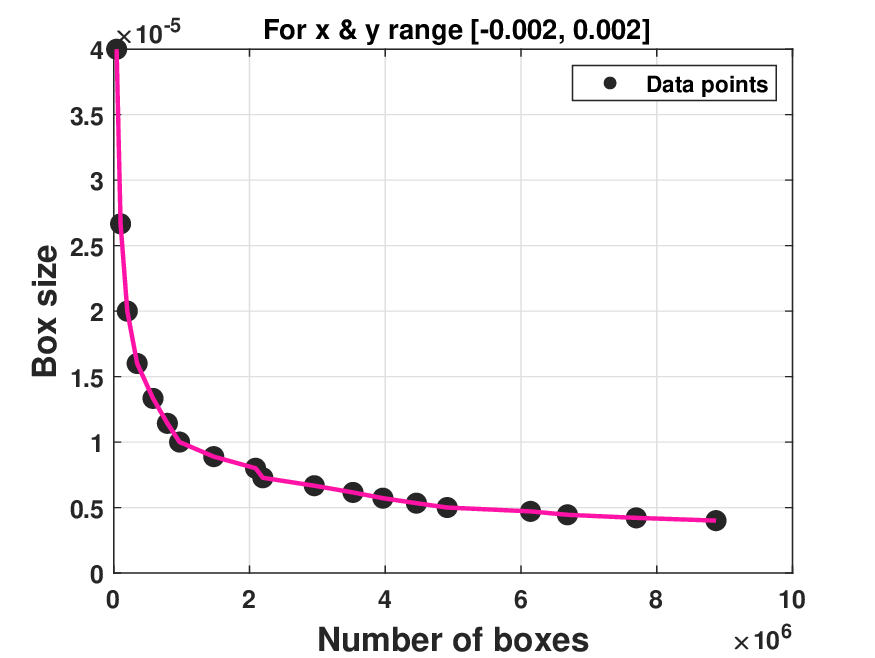}
        \caption{}
    \end{subfigure}
    \caption{Two dimensional plots of box size $(\varepsilon)$ versus number of boxes $(N(\varepsilon))$ required to cover the fractal surface corresponding to \figurename~\ref{f2} at three different zoom levels in the domain \( \xi \) and \( \eta \). Data points represent computed values from the box-counting process.}
    \label{f5}
\end{figure}

\begin{figure}[h]
    \centering
    \begin{subfigure}[b]{0.32\textwidth}
        \includegraphics[width=\linewidth]{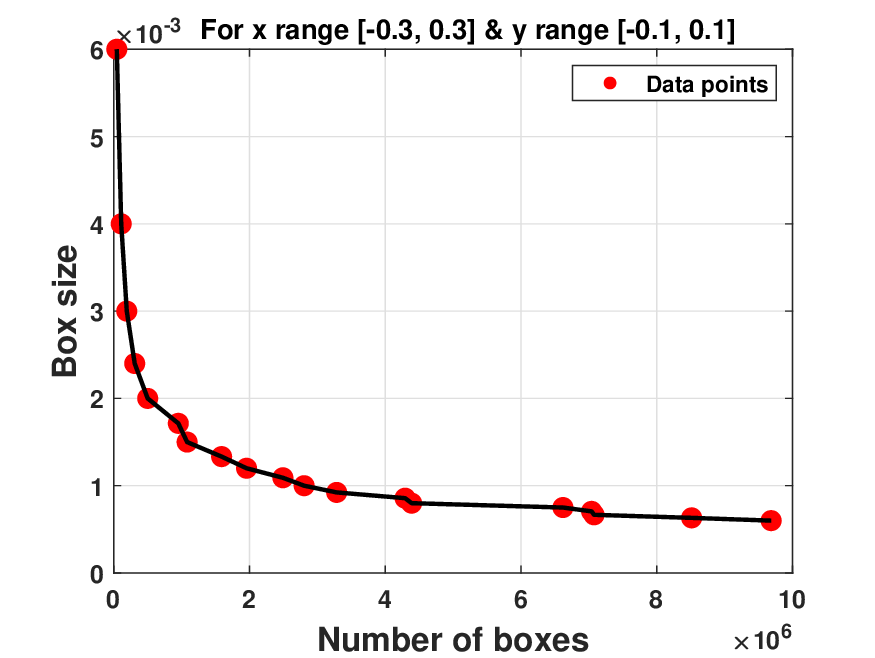}
        \caption{}
    \end{subfigure}
    \begin{subfigure}[b]{0.32\textwidth}
        \includegraphics[width=\linewidth]{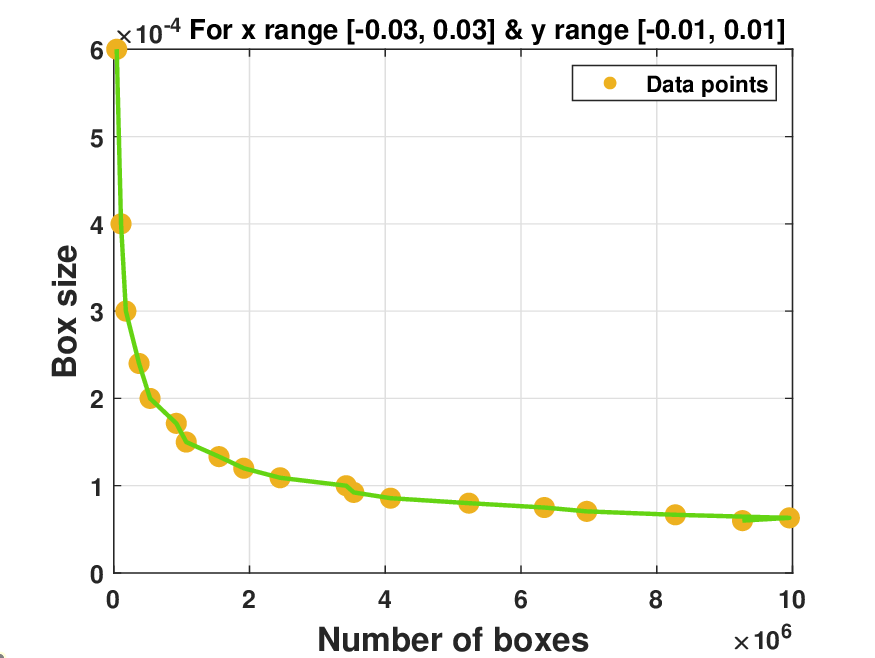}
        \caption{}
    \end{subfigure}
    \begin{subfigure}[b]{0.32\textwidth}
        \includegraphics[width=\linewidth]{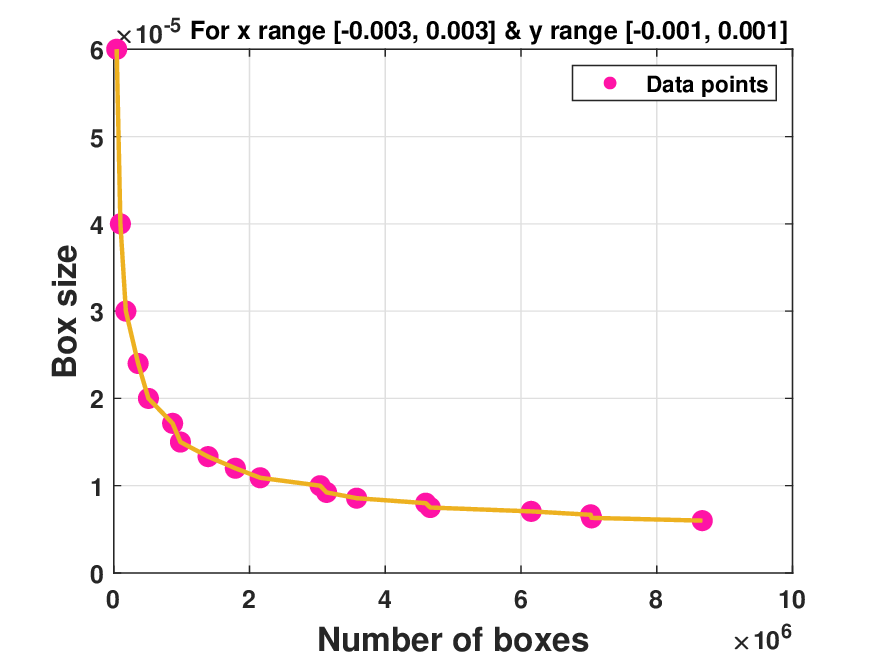}
        \caption{}
    \end{subfigure}
    \caption{Two dimensional plots of box size $(\varepsilon)$ versus number of boxes $(N(\varepsilon))$ required to cover the fractal surface corresponding to \figurename~\ref{f3} at three different zoom levels in the domain \( \xi \) and \( \eta \). Data points represent computed values from the box-counting process.}
    \label{f6}
\end{figure}

Figures~\ref{f4}--\ref{f6} illustrate the box-counting analysis performed for the self-similar structures generated by the breather interaction solutions of the KD equation. In each case, the variation of the number of occupied boxes $N(\varepsilon)$ with respect to the box size $\varepsilon$ is presented for different magnification levels of the corresponding structures.

It is observed that the number of covering boxes increases rapidly as the box size decreases, indicating the presence of increasingly fine geometrical details at smaller spatial scales. The nonlinear scaling behavior between $N(\varepsilon)$ and $\varepsilon$ reflects the multiscale complexity of the obtained patterns. The plotted data points correspond to the numerical values extracted during the box-counting computation and are used to estimate the associated fractal dimensions.

\FloatBarrier
\begin{table}[!htbp]
\centering
\caption{Fractal dimension estimates for \figurename~\ref{f1} obtained from voxel-based box-counting analysis. The quantities $RE$, $SE$, $\sigma_{\mathrm{boot}}$ and  $|\delta_D|$ denote the relative error, standard error, bootstrap standard deviation and  convergence measure, respectively.}
\begin{tabular}{ccccccc}
\hline
Figure  &\quad Dim. &\quad $RE$ &\quad $SE$ &\quad $\sigma_{\mathrm{boot}}$ &\quad $|\delta_D|$ \\
\hline
\figurename~\ref{f1a}   &\quad $2.41389$ &\quad $1.853\%$ &\quad $0.01338$ &\quad $0.02747$ &\quad $0.0010$ \\

\figurename~\ref{f1b}    &\quad $2.36674$ &\quad $3.309\%$ &\quad $0.02169$ &\quad $0.03693$ &\quad $0.0043$ \\

\figurename~\ref{f1c}  &\quad $2.42226$ &\quad $2.668\%$ &\quad $0.01727$ &\quad $0.03755$ &\quad $0.0073$ \\
\hline
\end{tabular}
\label{tab1}
\end{table}

\begin{table}[!htbp!]
\centering
\caption{Fractal dimension estimates for \figurename~\ref{f2} obtained from voxel-based box-counting analysis. The quantities $RE$, $SE$, $\sigma_{\mathrm{boot}}$ and  $|\delta_D|$ denote the relative error, standard error, bootstrap standard deviation and  convergence measure, respectively.}
\begin{tabular}{ccccccc}
\hline
Figure  &\quad Dim. &\quad $RE$ &\quad $SE$ &\quad $\sigma_{\mathrm{boot}}$ &\quad $|\delta_D|$ \\
\hline
\figurename~\ref{f2a}  &\quad $2.40359$ & \quad $1.574\%$ & \quad $0.01126$ &\quad $0.03344$ &\quad $0.0104$ \\

\figurename~\ref{f2b}  &\quad $2.39608$ &\quad  $1.123\%$ &\quad $0.00718$ &\quad $0.03740$ &\quad $0.0151$ \\
\figurename~\ref{f2c}   &\quad $2.33635$ &\quad $0.704\%$ &\quad $0.00984$ &\quad $0.02024$ &\quad $0.0018$ \\
\hline
\end{tabular}
\label{tab1}
\end{table}

\begin{table}[h!]
\centering
\caption{Fractal dimension estimates for \figurename~\ref{f3} obtained from voxel-based box-counting analysis. The quantities $RE$, $SE$, $\sigma_{\mathrm{boot}}$ and  $|\delta_D|$ denote the relative error, standard error, bootstrap standard deviation and  convergence measure, respectively.}
\begin{tabular}{ccccccc}
\hline
Figure  &\quad Dim. &\quad $RE$ &\quad $SE$ &\quad $\sigma_{\mathrm{boot}}$ &\quad $|\delta_D|$ \\
\hline
\figurename~\ref{f3a}  &\quad $2.37764$ &\quad $2.055\%$ &\quad $0.02526
$ &\quad $0.03390$ &\quad $0.0009$ \\

\figurename~\ref{f3b}   &\quad $2.41343$ &\quad $1.733\%$ &\quad $0.02331$ &\quad $0.03222$ &\quad $0.0128$ \\

\figurename~\ref{f3c}    &\quad $2.32855$ &\quad $0.225\%$ &\quad $0.01882$ &\quad $0.02948$ &\quad $0.0002$ \\

\hline
\end{tabular}
\label{tab1}
\end{table}
\FloatBarrier
The quantitative results of the box-counting analysis are presented in Tables 1--3. The estimated dimensions corresponding to all the examined structures are found to be non-integer, confirming the fractal nature of the breather interaction patterns generated by the KD system. In all cases, the dimensions lie approximately between $2.32$ and $2.44$, indicating significant geometrical complexity and multiscale irregularity in the obtained surface structures.

Table 1 presents the dimension estimates associated with Fig.~1. The estimated dimensions for Figs.~1(a)--1(c) are obtained as $2.41389$, $2.36674$ and  $2.42226$, respectively. These values remain consistently fractional under successive magnifications, demonstrating the persistence of self-similar scaling behavior. The corresponding relative errors are $1.853\%$, $3.309\%$ and  $2.668\%$, which remain sufficiently small and indicate reliable numerical estimation. Similarly, the standard error values $0.01338$, $0.02169$ and  $0.01727$ together with the bootstrap standard deviations $0.02747$, $0.03693$ and  $0.03755$ confirm the statistical stability of the inferred dimensions. Furthermore, the convergence measures $|\delta_D|=0.0010$, $0.0043$ and  $0.0073$ remain very small, showing that the calculated dimensions are numerically stable across different refinement levels.

The fractal dimension estimates corresponding to Fig.~2 are summarized in Table 2. The computed dimensions for Figs.~2(a)--2(c) are $2.40359$, $2.39608$ and  $2.33635$, respectively. These non-integer values again support the existence of fractal geometrical organization in the breather interaction profiles. The relative errors decrease from $1.574\%$ to $0.704\%$, indicating improved consistency for finer-scale structures. In addition, the standard errors remain low, namely $0.01126$, $0.00718$ and  $0.00984$, while the bootstrap deviations are found as $0.03344$, $0.03740$ and  $0.02024$. The corresponding convergence measures $0.0104$, $0.0151$ and  $0.0018$ further demonstrate that the estimated dimensions remain robust under successive refinement.

Table 3 contains the dimension analysis for the structures illustrated in Fig.~3. The estimated dimensions are obtained as $2.37764$, $2.41343$ and  $2.32855$ for Figs.~3(a)--3(c), respectively. These values again confirm the fractional scaling characteristics of the obtained patterns. The relative errors are calculated as $2.055\%$, $1.733\%$ and  $0.225\%$, showing an overall reduction in estimation uncertainty at finer resolutions. The associated standard errors are $0.02526$, $0.02331$ and  $0.01882$, whereas the bootstrap standard deviations are found to be $0.03390$, $0.03222$ and  $0.02948$. Moreover, the convergence measures $|\delta_D|=0.0009$, $0.0128$ and  $0.0002$ remain very small, providing additional evidence for the numerical consistency and reproducibility of the computed fractal dimensions.

Overall, the statistical indicators presented in Tables 1--3 demonstrate that the estimated fractal dimensions are both stable and reliable. The coexistence of non-integer dimensions, low relative errors, small standard errors and bootstrap deviations, together with strong convergence behavior, confirms that the observed self-similar structures are intrinsic features of the breather interaction solutions rather than numerical artifacts. These results provide quantitative evidence that nonlinear breather interactions in the $(2+1)$-dimensional KD equation naturally generate robust multiscale fractal geometries.

\section{Conclusions}\label{con} 

In this work, fractal structures arising from breather interactions in the $(2+1)$-dimensional Konopelchenko--Dubrovsky equation have been investigated through the Hirota bilinear method. Analytical breather interaction solutions were constructed and analysed under different parameter regimes in order to examine the spatial structures generated by the nonlinear interaction of the wave components.

The graphical analysis revealed that the obtained solutions develop highly intricate patterns exhibiting repeated structures under successive magnifications. Unlike earlier approaches, where fractal patterns were mainly generated through externally imposed auxiliary functions, the present study demonstrates that self-similar structures can emerge naturally from nonlinear breather interactions in a coupled nonlinear system. This suggests that coupled environments with strong nonlinear energy exchange may inherently support multiscale geometrical organization.

To quantitatively characterize the observed structures, a three-dimensional voxel-based box-counting analysis was performed. The estimated dimensions were found to be non-integer for all examined cases, lying approximately between $2.32$ and $2.44$, confirming the fractal nature of the generated patterns. In particular, the computed dimensions include values such as $2.41389$, $2.36674$, $2.42226$, $2.40359$, $2.39608$, $2.33635$, $2.37764$, $2.41343$ and  $2.32855$, indicating persistent multiscale complexity under successive magnifications. In addition, statistical diagnostics including relative error, standard error, bootstrap standard deviation and convergence analysis were incorporated to examine the reliability and stability of the computed dimensions. The relative errors remained small, ranging from $0.225\%$ to $3.309\%$, while the convergence measures $|\delta_D|$ also stayed sufficiently low throughout the refinement process. The obtained results showed small numerical errors and strong convergence behavior, indicating that the estimated fractal dimensions are  numerically stable, statistically reliable and reproducible.

Overall, the combined analytical construction, graphical visualization and statistical dimension analysis provide a systematic framework for studying multiscale structures generated by breather interactions in higher-dimensional coupled nonlinear systems. The present findings indicate that nonlinear breather dynamics can naturally generate complex self-similar geometries, which may be relevant for understanding energy localization, nonlinear wave modulation and multiscale pattern formation in fluid dynamics, plasma physics and nonlinear optical systems.

\vspace*{0.5cm}
\noindent{\large{\textbf{Statement and Declarations}}}\\
\textbf{Competing interests and fundings}\\
No funding was received to assist with the preparation of this manuscript. Also, the author has no financial or proprietary interests in any material discussed in this article.\\
\textbf{Data availability statement}\\
The author confirms that there is no associated data available for the above research work. Data sharing does not
apply to this article as no new data were created or analyzed in this study.\\
\textbf{Conflict of Interest}\\
The author has no conflict of interest  \\\\  
{\large{\textbf{Authors' Contributions}}}\\\\
\textbf{Snehalata Nasipuri: }  Investigation, Methodology, Software, Writing original draft.\\ \textbf{Prasanta Chatterjee:} Conceptualization, Visualization, Supervision.\\
\textbf{Saugata Dutta: } Investigation, Methodology, Software, Writing original draft. 

\section*{Acknowledgement} \noindent \textbf{Snehalata Nasipuri} is grateful to Sikkim Manipal Institute of Technology (SMIT) and Sikkim Manipal University (SMU) for providing a postdoctoral research fellowship under the T.M.A. Pai University Research Fund (Ref. No. SMU/URC/DoR/2026-17) and \textbf{Saugata Dutta} (NTA Ref. No. 211610066362) is deeply thankful
to the University Grants Commission (U.G.C.) of India for their financial support,
which has aided him to carry out this investigation.

%


\end{document}